\documentclass[twocolumn,showpacs,preprintnumbers,amsmath,amssymb,superscriptaddress]{revtex4}

\usepackage[dvips]{color,graphicx}

\usepackage{graphicx}

\begin{document}
\draft

\title{Temperature Dependent Polarity Reversal in Au/Nb:SrTiO$_3$ Schottky Junctions}

\author{T. Susaki,$^{1}$ Y. Kozuka,$^{1}$ Y. Tateyama,$^{1}$ and H. Y. Hwang}
\address{Department of Advanced Materials Science, University of Tokyo, Chiba 277-8561, Japan}
\address{Japan Science and Technology Agency, Kawaguchi, 332-0012, Japan}

\date{\today}



\begin{abstract}
We have observed temperature-dependent reversal of the rectifying polarity in Au/Nb:SrTiO$_3$ Schottky junctions. By simulating current-voltage characteristics we have found that the permittivity of SrTiO$_3$ near the interface exhibits temperature dependence opposite to that observed in the bulk, significantly reducing the barrier width. At low temperature, tunneling current dominates the junction transport due both to such barrier narrowing and to suppressed thermal excitations. The present results demonstrate that novel junction properties can be induced by the interface permittivity. 
\end{abstract}
\maketitle


Bulk SrTiO$_3$ is quite a unique material, starting from the dielectric constant: around 300 at room temperature, it grows to $\sim$ 20,000 at 4 K, driven by a ferroelectric instability frustrated by quantum fluctuations.\cite{Muller} 
Due to this instability, the permittivity is strongly nonlinear in electric field.\cite{Sawaguchi,Saifi}
Electron doping in SrTiO$_3$ rapidly induces high mobility carriers and the lowest density known bulk superconductor.\cite{Schooley} The lattice permittivity is thought to be a key aspect for screening in these electronic properties as well. In recent years, the polarizability of SrTiO$_3$ at the nanoscale has emerged as centrally important in understanding the electronic structure and charge distribution at oxide heterointerfaces and in superlattices. SrTiO$_3$ is currently used as a substrate, (super)-conductor, high-K dielectric layer, insulating tunnel barrier, and semiconductor. As artificial structures approach atomic dimensions, deviations from bulk behavior become increasingly dominant. For example, the dielectric dead layer in ultrathin SrTiO$_3$ capacitors has been found to arise from a short-length-scale lattice polarization, which compensates incomplete electrode screening of the depolarization field.\cite{Stengel} The $3d$ charge confinement of the LaTiO$_3$ layers sandwiched with SrTiO$_3$ layers is weakened due to the nanometer scale lattice screening;\cite{Ohtomo,Hamann,Okamoto} here again, nanometer-scale lattice screening is found to give a very important contribution.\cite{Hamann,Okamoto} Finally, this same issue arises in the electron gas at the LaAlO$_3$/SrTiO$_3$ interface \cite{Ohtomo2,Park} or SrTiO$_3$-based field effect transistor;\cite{Nakamura} nanometer scale dielectric properties are key to understanding the charge profile, for which little is experimentally known, particularly at low temperatures.

For these and many other reasons, isolating and probing the short-length-scale lattice response of SrTiO$_3$ to strong electric fields is widely needed. Because these are primarily interface phenomena involving the relative displacement of cations from oxygen (with small atomic number), direct imaging or refinement of the atomic positions theoretically deduced \cite{Stengel,Hamann,Okamoto,Park} is quite experimentally challenging. One approach would be to use electrical probes for which the local polarization dominates transport. For this purpose, a Schottky barrier formed between high work function metals and electron doped SrTiO$_3$ (Nb:SrTiO$_3$), although not suitable for imaging of atomic structures nor first principles calculations, is an ideal platform to probe the strong local nonlinear polarizability, which is directly manifest in the band-bending profile near the interface. Due to a change in the $local$ permittivity as a function of temperature, which is completely opposite to that observed in the bulk, the rectification polarity changes as a function of temperature. The present result demonstrates that the atomic scale properties at the interface, such as a strong reduction of the barrier width on a nanometer scale, can dominate macroscopic junction transport properties.


We fabricated Schottky junctions by contacting Au to  0.01 wt \% Nb-doped SrTiO$_3$(100) (Nb:SrTiO$_3$) substrate surface. 
First we annealed the Nb:SrTiO$_3$ substrate at 850 $^{\circ}$C under an oxygen partial pressure of
5 $\times$ 10$^{-4}$ Torr for 30 minutes to remove surface contamination. Such annealing under oxidizing conditions suppresses the reverse-bias leakage current.\cite{Shimizu,Suzuki} We evaporated Au \textit{in-situ} by resistive heating, and then evaporated Al on the back of the substrate to form an Ohmic contact. 
Forward bias was defined so that the positive voltage was applied to Au.

\begin{figure}
\begin{center}
\includegraphics[width=6.5cm]{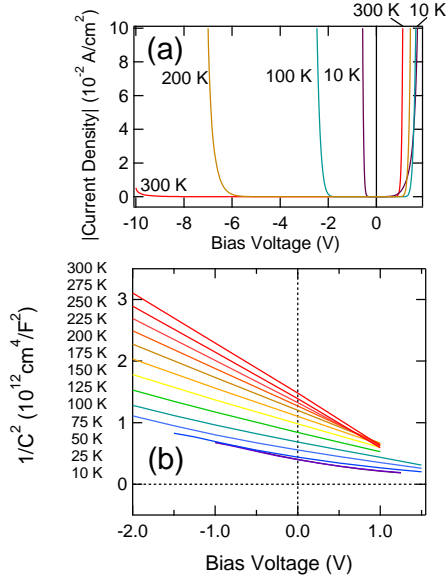}
     \caption{(Color online) (a)
     $I-V$ and (b) $C-V$ characteristics of a Au/Nb:SrTiO$_3$ Schottky junction at various temperatures. The junction area is $\sim$ 0.002 cm$^2$.}
\label{fig:linearIV}
\end{center}
\end{figure}

Figure~\ref{fig:linearIV} (a) shows current-voltage ($I-V$) characteristics of the Au/Nb:SrTiO$_3$(100) junction. We have found clear rectifying behavior at 300 K: the current steeply grows by applying a forward bias of $\sim$ 1 V, while almost no
reverse-bias current is observed down to $-$ 10 V. The sign of rectification is what is expected for a Schottky junction fabricated with an n-type semiconductor and a large work function metal. However, the absolute value of the bias voltage where the reverse-bias current starts to flow becomes smaller as the temperature is lowered. The characteristics are almost symmetric with respect to $V$ = 0 at 100 K, and finally the diode polarity becomes backward at 10 K.

We show the capacitance-voltage ($C-V$) characteristics measured at a frequency of 10 kHz in Fig.~\ref{fig:linearIV} (b). As already reported,\cite{Hayashi} 1/$C^2$ changes almost linearly as a function of the bias voltage at 300 K while it shows a U-like bending at low temperatures due to enhanced electric field dependence of the permittivity of SrTiO$_3$. By fitting 1/$C^2$ = $2 (V_{bi} - V)/(q \varepsilon_r \varepsilon_0 N_D)$ \cite{Sze} to the data at 300 K, the built-in potential $V_{bi}$ is estimated to be 1.76 V, assuming a relative permittivity of $\varepsilon_r$ = 300. Here $q$ is the elementary electric charge, $N_D$ is the donor density, and  $\varepsilon_0$ is the vacuum permittivity.

\begin{figure}
\begin{center}
\includegraphics[width=7.5cm]{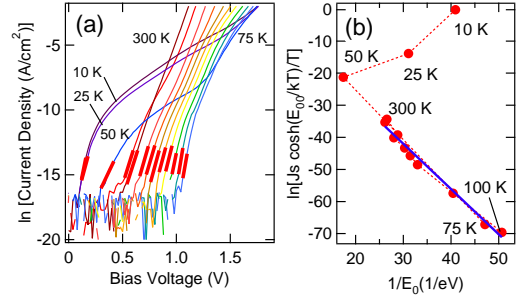}
     \caption{
     (Color online) (a) $I-V$ characteristics on a semilogarithmic scale between 300 K and 10 K. Linear fitting is shown by the bold lines. (b) $\ln[J_S\cosh(E_{\rm 00}/kT)/T] - 1/E_{\rm 0}$ plotted between 300 K and 10 K (dots connected with a dashed line). The bold line is a linear fit for the temperature region between 300 K and 100 K.
     }
\label{fig:semilog}
\end{center}
\end{figure}

We plot the forward-bias
$I-V$ characteristics on a semilogarithmic scale in Fig.~\ref{fig:semilog} (a). Between 300 K and $\sim$ 75 K the characteristics appear almost linear as expected for thermionic emission over the Schottky barrier.\cite{Sze} However, as the temperature is further lowered, a finite current starts to flow for bias voltage below 1 V, where no signal current is observed at 75 K. This low-temperature current does not show exponential voltage dependence, indicating that thermal processes do not play a dominant role below 75 K. 
Any increase in the junction current with decreasing temperature is quite unusual, implying significant temperature dependence of the Schottky barrier. Since the Schottky barrier height, which is given by the energy difference between the metal work function and semiconductor electron affinity, is expected to be independent of the temperature in Au/Nb:SrTiO$_3$, it must be the barrier width that significantly evolves with temperature. Because classical thermionic emission current does not depend on the barrier width but only on the barrier height, nonthermal transport, such as tunneling, dominates the temperature dependence in $I-V$ characteristics in the low-temperature regime.

In order to clarify the origin of the polarity reversal as well as the sharp transition in the junction transport mechanism around 100 K, we have modeled the effect of the temperature and electric-field dependence of the permittivity of SrTiO$_3$ on the $I-V$ characteristics. The crossover from low temperature direct tunneling to increased thermal contribution at higher temperature has been observed in Au/GaAs junctions.\cite{Padovani}  At finite temperatures, electrons are first thermally excited to an energy below the top of the barrier, and then tunnel through the remaining barrier in a process denoted thermally-assisted tunneling (thermionic field emission).  In this model, the linear dependence of ${\rm ln}[J_S {\rm cosh}(E_{\rm 00}/kT)/T]$ on $1/E_{\rm 0}$ shown in Fig.~\ref{fig:semilog} (b), where $E_0 \equiv q (\partial ln J/\partial V)^{-1}$ and $E_{\rm 00}$ = $E_{\rm 0}$(0 K)  $\propto [N_D/(m^* \varepsilon_r)]^{\frac{1}{2}}$ is approximated with the value at 10 K, indicates that the Schottky barrier profile does not  depend on the temperature.\cite{Padovani} Here, $J$ ($J_S$) is the (saturation) current density, $k$ is the Boltzmann constant and $m^*$ is the effective mass. The slope of this plot between 300 K and 100 K corresponds to a temperature independent barrier height of 1.41 V,  slightly smaller than $V_{bi}$ = 1.76 V found in the $C-V$ measurement at 300 K. 

\begin{figure}
\begin{center}
\includegraphics[width=7.5cm]{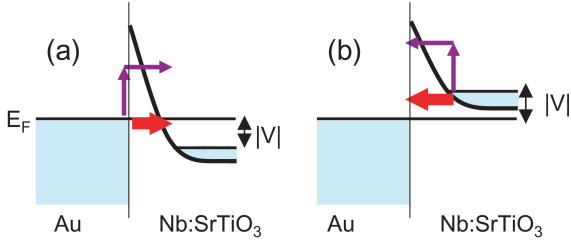}
     \caption{
     (Color online) Schematic potential profiles of the Au/Nb:SrTiO$_3$ Schottky junction under the reverse-bias condition (a) and forward-bias condition (b). The thick and thin arrows show the direct and thermally-assisted tunneling processes, respectively.}
\label{fig:schematic}
\end{center}
\end{figure}

For a given barrier profile $V(x)$, the thermally-assisted tunneling current density is calculated by \cite{Fonash}
\begin{eqnarray}
J = \int P(E_e,m^*)N(E_e)dE_e,
\label{eq:Fonash_J}
\end{eqnarray}
where the transmission probability $P(E_e,m^*)$ and "supply function" $N(E_e)$, which assures the occupancy/vacancy of the initial/final states, are expressed by
\begin{eqnarray}
P = \left[1 + \exp\left[-\frac{2i}{\hbar}\int \left[2m^* \left(E_e - V(x)\right) \right]^{1/2}dx  \right] \right]^{-1}
\label{eq:Fonash_P}
\end{eqnarray}
and
\begin{eqnarray}
N(E_e) = \left(A^* T/k\right) e^{\mu/kT} e^{-E_e/kT}\left(1-e^{-V/kT}\right),
\label{eq:Fonash_N}
\end{eqnarray}
where $E_e$ is the energy of electron, $\hbar$ is the Plank's constant divided by 2$\pi$, $A^*$ is the effective Richardson constant (156 A cm$^{-2}$K$^{-2}$)\cite{Sroubek} and $\mu$ is the chemical potential with respect to the conduction-band bottom in the semiconductor. As schematically shown in Fig.~\ref{fig:schematic}, direct and thermally-assisted tunneling is considered, while classical thermionic emission $over$ the barrier is neglected.  

Based on the form of 
\begin{eqnarray}
\varepsilon_r(E) = b/\sqrt{a + E^2}
\label{eq:field_dependent_epsilon}
\end{eqnarray}
for the electric-field ($E$) dependence of the relative permittivity of SrTiO$_3$,\cite{Suzuki}
where $a$ and $b$ are constants, the Schottky barrier potential $V(x)$ and relative permittivity $\varepsilon_r(x)$ of an n-type SrTiO$_3$/metal junction are analytically determined as a function of the distance from interface $x$ as \cite{Suzuki}
\begin{eqnarray}
- V(x) = \frac{\sqrt{a}b\varepsilon_0}{qN_D}\left\{\cosh\left[\frac{qN_D}{b\varepsilon_0}(W-x)\right] -1\right\} + V,
\label{eq:V_x}
\end{eqnarray}


\begin{eqnarray}
\varepsilon_r(x) = \frac{b}{\sqrt{a}\cosh\left[\frac{qN_D}{b\varepsilon_0}(W-x)\right]}
\label{eq:epsilon_x}
\end{eqnarray}

and

\begin{eqnarray}
W = \frac{b\varepsilon_0}{qN_D}\cosh^{-1}\left[1 + \frac{qN_D}{\sqrt{a}b\varepsilon_0}(- V + \phi_B)\right].
\label{eq:W}
\end{eqnarray}
Here, $W$ is the depletion layer width and $\phi_B$ is the barrier height. 
By fitting Eq.~(\ref{eq:field_dependent_epsilon}) to the permittivity measured in Al/SrTiO$_3$/Al capacitance structures, the temperature dependence of $a$ and $b$ has been deduced as $a(T) = [b(T)/\varepsilon_r(T,E=0)]^2$ and $b(T) = 1.37 \times 10^9 + 4.29 \times 10^7 T$ V/m, \cite{Yamamoto} where 
\begin{eqnarray}
\varepsilon_r(T, E=0) = \frac{1635}{\coth(\frac{44.1}{T})-0.937}
\label{eq:Barrett}
\end{eqnarray}
following Barrett's formula.\cite{Barrett} Then the temperature and electric-field dependence of $\varepsilon_r$ is uniquely given according to Eq.~(\ref{eq:field_dependent_epsilon}) and hence the barrier potential profile is determined by Eq.~(\ref{eq:V_x}) once the barrier height $\phi_B$ and the donor concentration are given.

\begin{figure}
\begin{center}
\includegraphics[width=6.5cm]{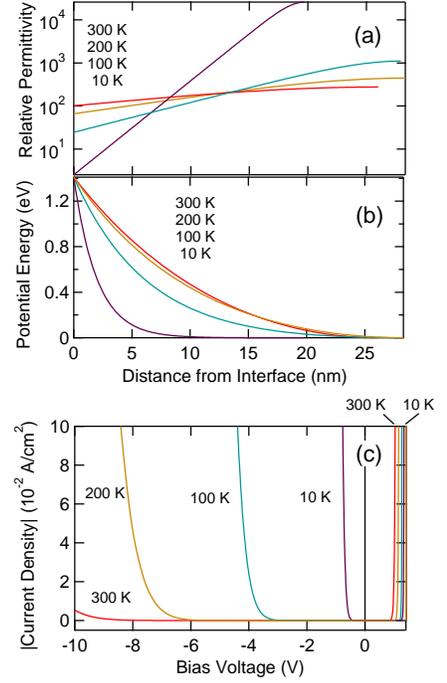}
     \caption{(Color online) (a) 
Simulated temperature dependence of the relative permittivity in the depletion region, (b) barrier potential profile, and (c) $I-V$ characteristics calculated for a barrier height of 1.41 V and a donor concentration of 5 $\times 10^{19}$ cm$^{-3}$.}
\label{fig:simulation}
\end{center}
\end{figure}

Using a constant barrier height of 1.41 V, we have found that a large temperature dependence in the reverse-bias current as well as the polarity reversal is reproduced, as shown in Fig.~\ref{fig:simulation} (c).  Panels (a) and (b) of the figure give the calculated temperature dependence of the permittivity and barrier profile. While the bulk permittivity of SrTiO$_3$ monotonically increases with decreasing temperature, the $local$ permittivity within $\sim$ 5 nm from the interface monotonically $decreases$ with decreasing temperature. Accordingly, the Schottky barrier width is reduced with decreasing temperature, giving a large tunneling current at lower temperature. The relatively small change both in the permittivity and in the barrier profile between 300 K and 100 K is consistent with the analysis in Fig.~\ref{fig:semilog} (b), where the barrier height, donor concentration and permittivity are assumed to be independent of the temperature.

Based on these simulated results, the diode polarity reversal can be understood straightforwardly.
 At low temperature, direct tunneling dominates due (i) to narrow barrier width corresponding to the suppressed local permittivity and (ii) to a negligible thermal contribution to the junction transport. In this regime, as reverse bias is applied, the number of initial states increases and correspondingly the junction current rapidly increases, as seen in Fig.~\ref{fig:schematic} (a). However, such increase in the current is not expected under forward biasing: once the biased conduction-band bottom becomes higher than the Fermi level of Au, the number of initial states does not change as a function of the bias voltage. 
On the other hand, the asymmetry in the number of initial states does not explicitly appear in the junction current at higher temperature: with increasing temperature (i) the barrier width is enlarged as the local permittivity recovers and (ii) the degree of thermal excitation increases, both of which lower the relative contribution of direct tunneling. 
We have further modeled the temperature variation of $N(E_e)$ and $V(x)$ independently, and have found that the effect of barrier narrowing  (temperature dependence in $V(x)$) dominates the temperature dependence in $I-V$ characteristics in the present case.

Despite these successes, this simple model has important limitations: We did not explicitly incorporate the short length scale variations of the permittivity ($\varepsilon(q)$), but used the bulk value ($q$ = 0)  to calculate the local $\varepsilon(x)$.
$I-V$ measurement probes the dielectric properties in the vicinity of the Au/Nb:SrTiO$_3$ interface, while $C-V$ measurement probes those at the "interface" at the edge of the depleted region. Although both techniques measure the local dielectric response, the deviation from bulk response is most severe at the Au/Nb:SrTiO$_3$ interface (the maximum internal electric field across the junction), which is reflected in $I-V$ characteristics. 
In actuality, given the extremely short distances over which the barrier varies (a few nanometers, consistent with the observation of direct tunneling \cite{Sroubek} and Fig.\ref{fig:simulation} (b)), the barrier should be calculated self-consistently in $q$. The absence of finite $q$ effects partly explains the large donor concentration required to quantitatively reproduce the experimental results. 
 Going further, at near unit cell length scales, the validity of a local effective description of the permittivity breaks down.\cite{Sawatzky} 
 
In closing, phenomena arising from the dramatic variations of  $\varepsilon(q,E,T)$ of SrTiO$_3$ at small length scales is crucial for many of the physical properties in ultra-thin film structures significantly dominated by interfaces. We have demonstrated that a simple metal Schottky junction can probe these effects, manifesting remarkable macroscopic features such as temperature dependent polarity reversal. Microscopically, these features originate from the strong local atomic displacements found to be generic in recent \textit{ab initio} studies of directly related structures, arising when the lattice positions are computationally relaxed. \cite{Stengel,Hamann,Okamoto,Park} This electric field induced polarization of the lattice should be a generic feature in the many experimental studies using SrTiO$_3$ in junctions, at interfaces, and in atomic-scale superlattices, giving unexpectedly strong internal electric fields and abrupt band-bending profiles.

We thank G. A. Sawatzky, N. Nagaosa, and A. Fujimori for useful discussions. The present work was partly supported by a Grant-in-Aid for Scientific Research (B) from the Ministry of Education, Culture, Sports, Science and Technology.


%


\end{document}